\documentclass[12pt]{article}
\usepackage{graphicx}
\usepackage{dcolumn}
\usepackage{bm}
\usepackage{graphics}
\usepackage{amsmath}
\usepackage{amssymb}
\usepackage{amscd}
\usepackage{afterpage}
\usepackage{float,times}
\usepackage{subfigure}
\usepackage{rotating}
\usepackage{multirow}
\usepackage{epsfig}
\usepackage{theorem}
\usepackage{moreverb}
\usepackage{euscript}
\usepackage{psfrag}

\begin{document}

\begin{titlepage}
\begin{center}
{\hbox to\hsize{\hfill March 2008 }}

\bigskip
\vspace{6\baselineskip}

{\Large \bf

(De)quantization of black hole charges  \\
}
\bigskip

\bigskip

{\bf 
Archil Kobakhidze and Bruce H. J. McKellar \\}
\smallskip

{ \small \it School of Physics, Research Centre for High Energy Physics, \\ The University of Melbourne, Victoria 3010, Australia\\}

\bigskip

\vspace*{.5cm}

{\bf Abstract}\\
\end{center}
\noindent
{\small 
We argue that  magnetic and electric charges of the Reissner-Nordstrom black hole are quantized in CP conserving theories. Dequantization phenomenon occurs when CP is broken either explicitly or effectively, and, as a result, pure magnetic black holes are not possible. Two examples illustrating this phenomenon are discussed.  One is the electric charge induced by the neutral pion field  in  a magnetically charged black hole, and another is the electric charge induced due to the massive spin-two field emerging from possible higher curvature terms.    }  

\bigskip

\bigskip

\bigskip

\end{titlepage}

\section{Introduction.} 

It is well-known that quantum mechanical consistency of a system an electron with charge $e$ and a monopole of charge  $g$ requires the quantization condition, 
\begin{equation}
eg=\frac{1}{2}n,~n Z~,
\label{1}
\end{equation} 
to be fulfilled \cite{Dirac:1931kp}.  The quantization condition (\ref{1}) has been generalized to a system of two dyons with charges $(q_e, q_m)$ and $(q_e', q_m')$ \cite{Schwinger:1966nj}, \cite{Zwanziger:1969by}:
\begin{equation}
q_eq_m' - q_e'q_m=\frac{1}{2}n,~n Z~.
\label{2}
\end{equation}   
For the charge-monopole system (e.g., $q_m=0$ and $q_e'=0$) (\ref{2}) reduces to (\ref{1}).  Consider now a dyon $(q_e, q_m)$ and its CP conjugated dyon $(-q_e, q_m)$. If the CP is an exact symmetry, then the quantization condition (\ref{2}) applied to the pair of such dyons implies the dyon charge quantization: 
\begin{equation}
q_e=\frac{1}{2}en,~n Z~,
\label{3}
\end{equation}   
where, according to (\ref{1}), we have taken an elementary magnetic charge to be $q_m=\frac{1}{2e}$. Thus dyon electric charges are either integer or half-integer in units of the elementary charge $e$. However,  in CP non-conserving theories electric charge of a dyon acquires a non-integral shift $\delta$, $q_e=\frac{e}{2}n+\delta$, which is proportional to the strength of the CP violation \cite{Witten:1979ey}. This means that magnetically charged particles necessarily carry non-zero electric charge $q_e=\delta$ (n=0) when the CP symmetry is broken.        

Monopoles \cite{'t Hooft:1974qc},\cite{Polyakov:1974ek} and dyons \cite{Julia:1975ff} emerge as a particle-like  solutions in spontaneously broken gauge theories. On the other hand  black holes also can carry electric and/or magnetic charges as well.  Does the charge quantization hold for such an objects?  The answer is yes. The necessity of black hole charge quantization can be understood from the following heuristic consideration.  As soon as we are interested in distantly separated  pair of dyonic black holes (with asymptotically flat space-time), they can be considered as usual point-like particles and the quantization condition (\ref{2}) must be applicable.   Also, the quantization condition (\ref{2}) is supported by the fact that  the electromagnetic angular momentum in black hole space-time has exactly the same form as in flat space-time \cite{Adler:1976zq}, together with the traditional semi-classical derivation of the condition (\ref{2}) in terms of the quantization of angular momentum. 

In this paper we argue that Dirac's quantization condition does indeed depend on the topology  rather than geometry of a black-hole space-time. Thus it does hold for charged black holes  as well as for dyonic systems in black-hole background space-times, as long as the electric and magnetic charges can be meaningfully defined (e.g., in static asymptotically flat space-times).  

It turns out that electric charge quantization holds only in CP conserving theories. Once CP symmetry  is broken the phenomenon of charge  dequantization  occurs. Witten showed that this was the case in flat space-time \cite{Witten:1979ey}, and we generalize the result to the space-time of black holes.  We discuss two further examples demonstrating this effect. One  involves a neutral pion field in the background of a magnetic black hole. Due to the axial anomaly, the pion field  couples topologically to two photons. While this interaction term does not  contribute to the energy momentum tensor, it does affect the dynamics of the pion field so that the pion carries classical hair outside the black hole horizon which in principle is detectable. At spatial infinity the pion field approaches its non-trivial vaccum expectation, and thus the anomalous pion-photon interaction term turns into an effective CP-violating $\theta$-term. The later induces an electric charge on the black which is non-integer multiple of the elementary charge $e$ in general.    

Another example illustrates the effect  of charge dequantization within the theory with higher curvature terms. The higher curvature theories contain a massive spin-two field. According to the no-hair theorem \cite{Bekenstein:1996pn},  this spin-2 field vanishes outside the (Schwarzchild) black hole horizon, however its longitudinal part carries quantum hair, as discovered in \cite{Dvali:2006az}. Assuming that the field providing  the quantum hair couples to the photon through the topological interaction term, we again generate a non-integer electric charge on the black hole. Obviously, the charge (de)quantization phenomenon  can be applied to any other U(1) charge which is supported by the gauge field.     

The effect of charge (de)quantization discussed in this paper can be contrasted with recent observation on the quantization of global charges \cite{Dvali:2007hz}.   Dvali argued that a global charge, i.e. the charge which is not coupled to a long-range (massless) field,  is necessarily quantized with maximal periodicity $n_{\rm max}=[M_P^2/m^2]$. The origin of this charge quantization (see \cite{Dvali:2007hz} for details) is entirely different from the one applying to fields which couple to massless fields, which we discuss in this paper.  Note also, that gauge theories with both electric ($g$) and magnetic ($1/g$) charges do not have a sensible global limit, $g\to 0$, and thus our result on dequantization is not in contradiction with the quantization phenomenon discussed in \cite{Dvali:2007hz}.

\section{Dyonic Reissner-Nordstrom black hole.}

Let us consider a Reissner-Nordstrom black hole with electric and magnetic charges. It is described by the metric tensor
\begin{equation}
g_{\mu\nu}={\rm diag}\left[A,~-A^{-1},~ -r^2,~-r^2\sin^{2}\theta  \right]~,
\label{4}
\end{equation}
where $A=1-G_N\left(\frac{2M}{r}-\frac{q^2_e}{r^2}-\frac{q^2_m}{r^2}\right)$, and by the electromagnetic field strength two-form, 
\begin{equation}
F=-\frac{q_e}{r^2}dt\wedge dr-q_m\sin\theta d\theta\wedge d\phi~.
\label{5}
\end{equation}
 The dual two-form, $*F=\frac{\sqrt{-g}}{2}\epsilon_{\mu\nu\rho\sigma}F^{\rho\sigma}dx^{\mu}\wedge dx^{\nu}$, reads:
\begin{equation}
*F=-\frac{q_m}{r^2}dt\wedge dr+q_e\sin\theta d\theta\wedge d\phi
\label{6}
\end{equation}
Note that the above two-forms satisfy the source-free equation of motion $d*F=0$.  Electric-magnetic duality is manifest in the above equations, since $*F$ is obtained from $F$ by  the transformation $q_e\to q_m$, $q_m\to - q_e$. 

The metric (\ref{4}) has a physical singularity at the origin $r=0$, and thus topologically space-time is $R^2\times S^2$. On the other hand (since the space-time is assymptotically flat), we can define global electric
\begin{equation}
q_e=\frac{1}{4\pi}\int_{S^2_{\infty}} *F,
\label{7}
\end{equation}
 and magnetic 
 \begin{equation}
 q_m=\frac{1}{4\pi}\int_{S^2_{\infty}}F
 \label{8}
 \end{equation}
 charges as the total electric and magnetic fluxes through the two-shpere at infinity.  
 
In quantum theory the primary field is the vector potential one-form $A=A_{\mu}dx^{\mu}$, and the two-form is given as $F=dA$. Thus alongside with the equation of motion $d*F=0$ we have Bianchi identity, $dF=0$. The duality symmetry allows us to alternatively define a dual one-form, $\tilde A$ from $*F=d\tilde A$, as a primary field. With the later choice the equation of motion and the Bianchi identity are interchanged, but both equations still hold.  
 
 In the original derivation of the quantization condition (\ref{1}) Dirac made use of  the singular  gauge potential. The equation (\ref{1}) readily follows from the condition of non-observability of this singularity (Dirac's string). Indeed the Dirac string is known to be a gauge  artifact, and the appropriate gauge exists where the potential is regular, but multiply defined object \cite{Ignatiev:1998jq}. Thus the "true" nature of Dirac's quantization is encoded in the non-trivial topology, namely $R^1\times R^3/\{0\}$, which is diffeomorphic to $R^2\times S^2$. But this is exactly the topology of black hole space-times. This simple topological argument   again leads to the conclusion that Dirac's quantization must hold in black-hole space-times (contrary to some claims in the literature, see e.g. \cite{Ignatiev:1998jq}).   
 
 Indeed, since the electric charges in  (\ref{7}) and (\ref{8}) are defined on  a topological $S^2$, we are primarily interested in induced gauge potentials and gauge transformations on a two-shpere. Denote ${\cal U}$ the space of all such restricted gauge potentials, $A^{g}=A+g^{-1}dg$, and by ${\cal G}$, the space of all continuous gauge transformations $g$ on $S^2$.   Then the restricted gauge orbit space  is the quotent ${\cal U}/{\cal G}$. Because of ${\cal U}$ is topologically trivial, i.e. $\pi_n({\cal U})=0$ for any $n$,  fron the exact homotopy sequence one obtains
 \begin{equation}
 \pi_2({\cal U}/{\cal G})=\pi_1({\cal G}) = Z~
 \label{9}
 \end{equation}   
 for $G=U(1)$ gauge group. This simply means that the quantum mechanical system of an elementary charge  $e$ moving in the field of magnetically charged black hole is consistent once the quantization condition 
 \begin{equation}
 q_me=\frac{1}{2}n
 \label{10}
 \end{equation}
 is satisfied.
Analogously, for the quantum mechanical system of a monopole of charge $g$ and an electrically charged black hole, consistency demands the quantization condition
\begin{equation}
q_eg=\frac{1}{2}n
\label{11}
\end{equation} 
In the presence of both an elementary charge and an elementary monopole we can use Dirac's quantization condition to obtain the elementary magnetic charge $g=\frac{1}{2e}$. Then (\ref{11}) becomes
\begin{equation}
q_e= n e
\label{12}
\end{equation}
This is the quantization condition for the electric charge of the black holes in terms of "elementary" charge $e$. 

As we have mentioned above a charge dequantization phenomenon \cite{Witten:1979ey} occurs in the case of CP violation. This can be most easily seen within the model where the CP violation is provided by a Chern-Simons term, i.e.
\begin{equation}
{\cal L}_{CS}=\frac{\theta e^2}{8\pi}F\wedge F~. 
\label{13}
\end{equation}
Although the above term is a boundary term and thus does not affect the dynamics, it does contribute to the electric charge (\ref{7}) of the black hole, which now reads 
\begin{equation}
Q_e=\frac{1}{4\pi}\int_{S^2_{\infty}}\left( *F+\frac{\theta e^2}{2\pi}F \right )
\label{14}
\end{equation}
The arguments which lead to equation (\ref{12}) are now applied to (\ref{14}), and thus we obtain
\begin{equation}
q_e=\left ( n-\frac{m\theta}{4\pi}\right)e~,
\label{15}
\end{equation} 
where we have used $q_m=\frac{m}{2e}$. The charge dequantization (\ref{15}) implies the {\it nonexistence of  pure magnetic black holes in CP non-conserving theories}.

\section{Induced charge on the black hole}

Interestingly, the phenomenon of dequantization (\ref{15}) can effectively emerge in some theories even in the absence of a Chern-Simons term (\ref{13}). Below we discuss some two examples. 

\paragraph{Electric charge induced by the neutral pion} The no-hair theorem (conjecture actually) \cite{Bekenstein:1996pn} is one of the most important results in black hole physics. According to this theorem, the only dynamical field strengths present in exterior of (stationary and asymptotically flat) black hole solutions are those required by a local gauge invariance. For example, the gauge invariance of electrodynamics coupled to gravity implies non-vanishing of electric and/or magnetic fields around the black hole. On the contrary, the fields  which are not related to a local gauge invariance can not be observed outside the black hole horizon.   For example, the baryonic charge does not couple to a gauge field, and thus it can not be measured by the exterior observer. This implies that the black hole made of baryons and the one made of anti-baryons are indistinguishable as long as they have the same electric charge.  We will see below that in ceratain cases a global baryonic charge can induce an electric charge which can be detected by the corresponding electrostatic field. 

Baryons and mesons are described reasonably well by effective chiral Lagrangians. We consider a gauged two-flavour $SU(2)_L\times SU(2)_R$ effective chiral theory. Upon the chiral symmetry breaking the low energy degrees of freedom (pions) are combined in the unitary matrix field  $U=\exp{(i\pi(x)^a\tau^a)}$  which belongs to a quotent  $SU(2)_L\times SU(2)_R/SU(2)_V$ [here $\tau^a$ are the Pauli matrices, such that $\rm Tr(\tau^a\tau^b)=2\delta^{ab}$ ].
\begin{eqnarray}
\frac{1}{\sqrt{-g}}{\cal L}_{\rm chiral}={\cal L}_{\rm pion}+{\cal L}_{\rm anom} \\  \nonumber \\
{\cal L}_{{\rm pion}}= +\frac{f_{\pi}^2}{16}g^{\mu\nu}{\rm Tr}\left(D_{\mu}UD_{\nu}U^{\dagger}\right)+\frac{(f_{\pi}m_{\pi})^2}{16\hat m}{\rm Tr}\left ( M_q[U+U^{\dagger}-2]\right)~,\\ \nonumber \\
{\cal L}_{\rm anom}=\frac{eN_c}{48\pi^2}\frac{\epsilon^{\mu\nu\rho\sigma}}{\sqrt{-g}}A_{\mu}{\rm Tr}\left [\hat Q
 \left(\partial_{\nu}U\partial_{\rho}U^{\dagger}\partial_{\sigma}U U^{\dagger}+ U^{\dagger}\partial_{\nu}U\partial_{\rho}U^{\dagger}\partial_{\sigma}U \right)\right] \nonumber \\
 +\frac{ie^2N_c}{48\pi^2}\frac{\epsilon^{\mu\nu\rho\sigma}}{\sqrt{-g}}A_{\mu}F_{\nu\rho}{\rm Tr}\left[2 \hat Q^2\{\partial_{\sigma}U, U^{\dagger}\}+\hat Q\{\partial_{\sigma}UU^{\dagger}, U\hat Q U^{\dagger}\}  \right ]
\label{16}
\end{eqnarray}
The various contributions in (16) are described as follows. The first term in (17) is the $SU(2)_L\times SU(2)_R$ invariant term for the pion fields. $D_{\mu}=\partial_{\mu}-ieA_{\mu}[\hat Q,.]$ is the hermitian covariant derivative with respect to electromagnetic gauge symmetry, with $A_{\mu}$ being the  photon field and $\hat Q$ the charge matrix, $\hat Q\equiv \frac{1}{2}\tau^3+\frac{1}{6}{\bf 1}={\rm diag}[2/3, -1/3]$.  The second term in (17) describes the mass of the pion fields which emerge as a result of explicit breaking of chiral symmetry by the current quark masses, $M_q={\rm diag}[m_u,m_d]$. $f_{\pi}$ and $m_{\pi}$ are the pion decay constant and the pion mass respectively, and $\hat{m}=(m_u+m_d)/2$ is the mean quark mass. The part of the total Lagrangian in (18) is the effective term coming from the $U(1)_A$ chiral anomaly \cite{Callan:1983nx}, where $N_c=3$ is the number of colours.    It is easy to see that (18) correctly describes anomalous processes such as $\pi^0\to\gamma \gamma$ or the anomalous $\gamma \pi^{+}\pi^{-}\pi^{0}$ vertex.     

Here we are interested in excitations of a neutral pion field\footnote{Dvali has pointed out to us that the effect discussed here can be understood in terms of a topologically massive two-form field dual to the pion field (see e,g, \cite{Dvali:2005an} for the dual description of the pseudoscalar axion field). }, 
\begin{equation}
U(x)=\exp{\left( i\pi(x)\tau^3 \right)}~,
\label{17}
\end{equation}
in the background of a magnetically charged black hole.  The Lagrangian (16) takes the form:
\begin{equation}
\frac{1}{\sqrt{-g}}{\cal L}_{\rm chiral}=\frac{f_{\pi}^2}{8}g^{\mu\nu}\partial_{\mu}\pi(x)\partial_{\nu}\pi(x)+
\frac{(f_{\pi}m_{\pi})^2}{4}[\cos(\pi(x))-1]-\frac{e^2N_c}{24\pi^2}\frac{\epsilon^{\mu\nu\rho\sigma}}{\sqrt{-g}}\pi(x)F_{\mu\nu}F_{\rho\sigma}~.
\label{18}
\end{equation} 
Since $\pi(x)$  is a pseudoscalar field, the above theory is evidently CP conserving.  However, the boundary condition $\lim_{x\to\infty}|U(x)|=1$ implies $\lim_{x\to \infty}\pi(x)=\pi n$, and thus the last term in (\ref{18}) turns into the Chern-Simons term (\ref{13}) with $\theta=-\frac{N_c}{3}n$. Indeed, it is easy to see that spherically symmetric solution for the pion field goes as $\pi(r)=\pi n+{\cal O}(1/r)$. Thus,  according to (\ref{15}), the neutral pion field does induce an electric charge, $q_e=+\frac{n}{4\pi}e$,  on the black hole [note that we have set $N_c=3$ here]. Interestingly, in the context of the gauged Skyrme model, the effective $\theta$ is ultimately related to the baryon number \cite{Callan:1983nx},
\begin{equation}
n_B=\frac{eq_m}{2\pi}\left (\pi(\infty)-\pi(r_+)\right )~,
\label{19}
\end{equation} 
and the value of the pion field at the event horizon $r=r_+$, so that
\begin{equation}
q_e=\frac{n}{2\pi q_m}\left(n_B+\frac{eq_m}{2\pi}\pi(r_+)\right )~.
\label{19a}
\end{equation} 

\paragraph{Charge induced by  higher spin fields} 
As a second example of induced charge on the black hole we consider the model with curvature terms of a higher degree. For simplicity we consider here only quadratic terms in the curvature. The most general gravitational Lagrangian can be written as:
\begin{equation}
\frac{1}{\sqrt{-g}}{\cal L}_{\rm high~ cur}=aR^2+bR_{\mu\nu}R^{\mu\nu}~.
\label{20}
\end{equation}
The third possible invariant $R_{\mu\nu\rho\sigma}R^{\mu\nu\rho\sigma}$ is removed using the Gauss-Bonnett theorem. We also assume the specific relation $4a+b=0$ between the dimensionless couplings. With this relation we remove an extra scalar degree of freedom which is not essential for our discussion here. Then the Lagrangian (\ref{20}) can be rewritten in an equivalent form expressed in term of a massive spin two field\footnote{The massive spin-two field $\pi_{\mu\nu}(x)$ mixes with usual gravitons of Einstein  theory. The physical massive spin-two field has a wrong sign kinetic term, i.e. it is the ghost field. This issue is not important for our discussion in what follows.} $\pi_{\mu\nu}(x)$,
\begin{equation}
\frac{1}{\sqrt{-g}}{\cal L}_{\rm high~cur}=-\frac{1}{4b}\pi_{\mu\nu}\pi^{\mu\nu}+\pi_{\mu\nu}\Pi^{\mu\nu}~,
\label{21}
\end{equation} 
 where $\Pi_{\mu\nu}=R_{\mu\nu}-\frac{1}{4}g_{\mu\nu}R$. Using the equation of motion,
 \begin{equation}
 \pi_{\mu\nu}=2b\Pi_{\mu\nu}~,
 \label{22}
 \end{equation}
one can easily check that (\ref{20}) and (\ref{21}) are indeed equivalent. In the background of Schwarzchild black hole $R_{\mu\nu}=0$, and thus we have the trivial solution $\pi_{\mu\nu}(x)=0$. This solution reflects the well-known fact that quadratic in curvature terms do not affect the Schwarzchild black hole solution.  

Recently, it has been demonstrated in \cite{Dvali:2006az}  that massive higher spin fields, and in particular massive spin-two fields, might carry quantum hair in the black hole background. This type of hair is detectable in quantum mechanical Aharonov-Bohm-type of experiments, providing a  certain topological interaction of the longitudinal part of the spin-two field and a string is present. We decompose $\pi_{\mu\nu}(x)$ into a transverse components $\tilde \pi_{\mu\nu}(x)$ (contains two physical states) and the components carrying three massive longitudinal degrees $B_{\mu}(x)$,
\begin{equation}
\pi_{\mu\nu}=\tilde \pi_{\mu\nu}+\partial_{\mu}B_{\nu}+\partial_{\nu}B_{\mu}~,
\label{23}
\end{equation}
and allow the coupling \cite{Dvali:2006az}, 
\begin{equation}
\alpha \epsilon^{\mu\nu\rho\sigma}B_{\mu\nu}F_{\rho\sigma}~,
\label{24}
\end{equation}  
where, $\alpha$ is some constant, $B_{\mu\nu}=\partial_{\mu}B_{\nu}-\partial_{\nu}B_{\mu}$ and $F_{\mu\nu}$ is the electromagnetic field strength. Note that the reason why we can write this interaction is that (\ref{24}) is topological and does not affect the dynamics. It is perfectly consistent with the trivial solution $\pi_{\mu\nu}=0$ to have a nontrivial magnetic configuration for the two-form field $B=B_{\mu\nu}dx^{\mu}\wedge dx^{\nu}$
\begin{equation}
B=-q_m\sin\theta d\theta \wedge d\phi~.
\label{25}
\end{equation}
It is clear now that because of (\ref{25}) and the topological interaction (\ref{24}) an electric charge $q_{e}=4\alpha q_m$ will be induced on a black hole. Note that, $q_m$ in this case is not of electromagnetic origin. Thus starting from uncharged Schwarzchild black hole one arrives to the conclusion that it has a charge $q_e$ which can be detected by measuring  the local electrostatic field around the black hole. 

\section{Conclusion}

In this paper we have argued that electric and magnetic charges of a dyonic black hole are quantized in CP conserving theories. If CP is broken the electric charge gets non-integer contribution proportional to the strength of the CP violation. In this case pure magnetic black holes do not exist. We also demonstrate that charge dequantization can be induced by the neutral pion field. Finally, the electric charge can be induced by higher spin fields on an uncharged black hole.  In particular, this can happen in a theory with higher curvature terms, providing the topological interactions (\ref{24}) are present.   We find these effects rather amusing.

\subparagraph{Acknowledgments.}   We would like to thank Gia Dvali for valuable comments. The work was supported by the Australian Research Council. 


\end{document}